# In-situ alignment of anisotropic hard magnets of 3D printed magnets

M. Suppan[1], C. Huber[1], K. Mathauer[1], C. Abert[1,2], F. Brucker[1], J. Gonzalez-Gutierrez[3,7], S. Schuschnigg[3], M. Groenefeld[4], I. Teliban[4], S. Kobe[5], B. Saje[6], D. Suess[1,2]

[1] Faculty of Physics, University of Vienna, 1090 Vienna, Austria
[2] Platform MMM Mathematics–Magnetism–Materials, University of Vienna, 1090 Vienna, Austria
[3] Institute of Polymer Processing, Montanuniversitaet Leoben, 8700 Leoben, Austria
[4] Magnetfabrik Bonn GmbH, 53119 Bonn, Germany
[5] Dep. of Nanostructured Materials, Jožef Stefan Institute, 1000 Ljubljana, Slovenia
[6] Kolektor Magnet Technology GmbH, 45356 Essen, Germany
[7] Luxembourg Institute of Science and Technology, 4362 Esch-sur-Alzette, Luxembourg

dieter.suess@univie.ac.at

**Abstract**

Within this work, we demonstrate in-situ easy-axis alignment of single-crystal magnetic particles inside a polymer matrix using fused filament fabrication. Two different magnetic materials are investigated: (i) Strontium hexaferrite inside a PA6 matrix, fill grade: 49 vol% and (ii) Samarium iron nitride inside a PA12 matrix, fill grade: 44 vol%. In the presence of the external alignment field, the strontium hexaferrite particles inside the PA6 matrix can be well aligned with a ratio of remanent magnetization to saturation magnetization of 0,70. No significant alignment for samarium iron nitride could be achieved. The results show the feasibility to fabricate magnets with arbitrary and locally defined easy axis using fused filament fabrication since the permanent magnets used for the alignment (or alternatively an electromagnet) can be mounted on a rotatable platform.

## Introduction

3D printing of permanent magnets has recently received considerable attention[1,2]. In previous work, the realization of isotropic magnets was demonstrated using fused filament fabrication, which results in the fabrication of polymer-bonded magnets, with properties very similar to state-of-the-art magnets fabricated by injection molding, where isotropic magnets were produced[1]. The fused filament fabrication (FFF) method is a simple and fast manufacturing option for small objects. It is an additive 3D-printing technology that uses economical, commercially available FFF 3D-printers. A significant step forward is the fabrication of anisotropic magnets, where each magnetic particle is composed of a single magnetic crystal with a single easy axis. In FFF printing this can be achieved by applying an external field during printing. If an external field $H_{ext}$ is applied, a mechanical torque $T$ acts on the particle. For a given external field $H_{ext}$ the mechanical torque can be calculated by first determining the micromagnetic equilibrium magnetization $M(x)$ within each particle. Then the mechanical torque $T$ can be calculated by

$$\mathbf{T} = \mu_0 \int_{V_{particle}} \mathbf{M}(\mathbf{x}) \times \mathbf{H}_{ext} dV, \qquad (1.1)$$

where $V_{particle}$ is the volume of the particle. The equilibrium direction of $M(x)$ depends on crystalline anisotropy, demagnetizing energy, which also gives rise to shape anisotropy, the exchange energy and of course on the external field $H_{ext}$. For single domain particles without shape anisotropy the mechanical torque tries to align the easy axis parallel to the external field $H_{ext}$. The counter acting torque is due to the viscosity and friction of molten compound. An alignment model that describes coupled particle-fluid-magnetic field interactions during additive manufacturing of anisotropic bonded magnets is presented by Sarker et al. [3].

Experimentally, alignment of fused filament was realized by Sonnleitner et al. [4] by mounting strong permanent magnets below the building platform.. Here the magnetic stray field of a permanent magnet aligns the particles during the printing process. An alignment after the fabrication process is reported by Gandha et al. [5]

This paper presents a significant step forward by producing magnets where the easy axis can be locally aligned in-situ during the printing process. This is realized by redesigning the printer nozzle of a fused filament fabrication printer (FFF) by placing permanent magnets next to the nozzle. In principle, this set-up of the permanent magnets can be realized on a rotational platform so that the magnetic field can be rotated during the printing process. Alternatively, electromagnets could be used to realize the required fields for the alignment[6].

The magnetic stray field of the permanent magnets is optimized so that it aligns the easy axis of anisotropic ferromagnetic particles inside a paste-like compound material when it is in the molten state.

## Results

The bonded magnets are produced with a Velleman's K8200 printer and an E3D's Titan Aero extruder. The filament consists of a compound of magnetic powders and the binder. two different binders are used, in particular:

- Strontium-hexaferrite ($SrFe_{12}O_{19}$) inside a PA6 matrix (Sprox® 10/20p), fill grade: 49 vol%
- Samarium-iron-nitride ($Sm_2Fe_{17}N_3$) inside a PA12 matrix, fill grade: 44 vol%

The Sprox® compound is prefabricated by Magnetfabrik Bonn. This strontium hexaferrite powder consists of cuboid flakes with approximate dimensions in the range 6x2x2 µm³. The $Sm_2Fe_{17}N_3$ particles are spherical with a diameter in the range of $2-4$ µm. The magnetic filaments were produced at the Institute for Polymer Processing by a screw extruder [7]. The Sprox® compound is prefabricated by Magnetfabrik Bonn. This strontium hexaferrite powder consists of cuboid flakes with approximate dimensions in the range of 6x2x2 µm³ with $B_r$=196 mT, and with $H_{cJ}$=183 kA/m. The $Sm_2Fe_{17}N_3$ particles are spherical with a diameter in the range of $2-4$ µm with $B_r$ =1.31 T and $H_{cJ}$ = 889 kA/m
In order to realize the alignment of the magnetic particles, two cylindrical $Sm_2Co_{17}$ permanent magnets with a diameter of 12 mm and a height of 18 mm are placed next to the nozzle. The magnetic field generated from the magnets aligns the magnetic particles of the molton filament within the nozzle when the polymer is in the liquid state. The magnetic properties of the used $Sm_2Co_{17}$ permanent magnets are summarized in Table 1.

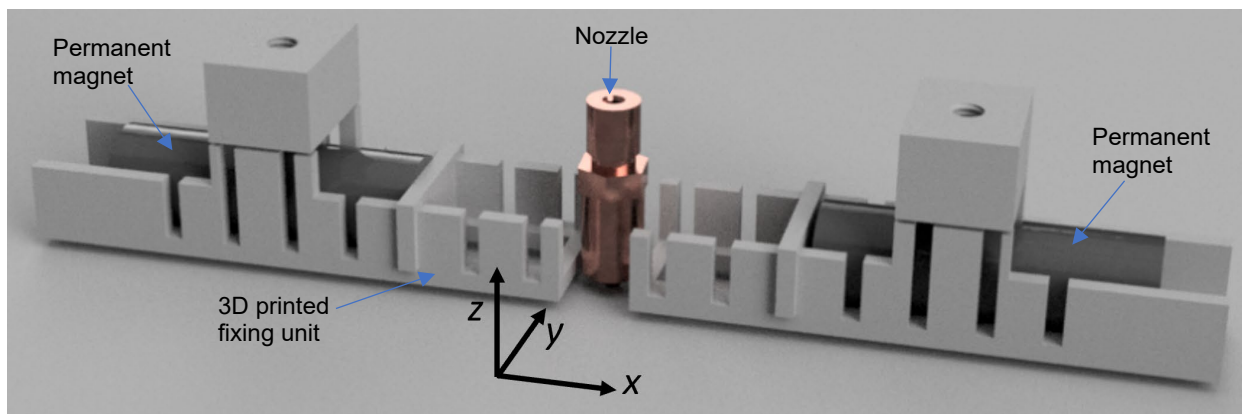

*Figure 1: Flexible 3d printed fixing unit for the bias magnets that is placed next to the nozzle.*

Detailed optimization of the field strength during the printing of the magnet is essential. Too small fields cannot align the particles, too large fields attract due to the larg fields gradients the magnetic particles towards the poles of the permanent magnets, and no printed object can be realized. To be able to adjust the field, a flexible fixing unit for the magnets was printed from polylactides, as shown in Fig. 1. Depending on the required field strength, the distance between the two cylindrical magnets was adjusted. In Fig. 2, the modification of the 3D-Printer is illustrated. The printer has a spacious scaffold out of aluminum, which provides sufficient space to attach the fixing units around the printhead.

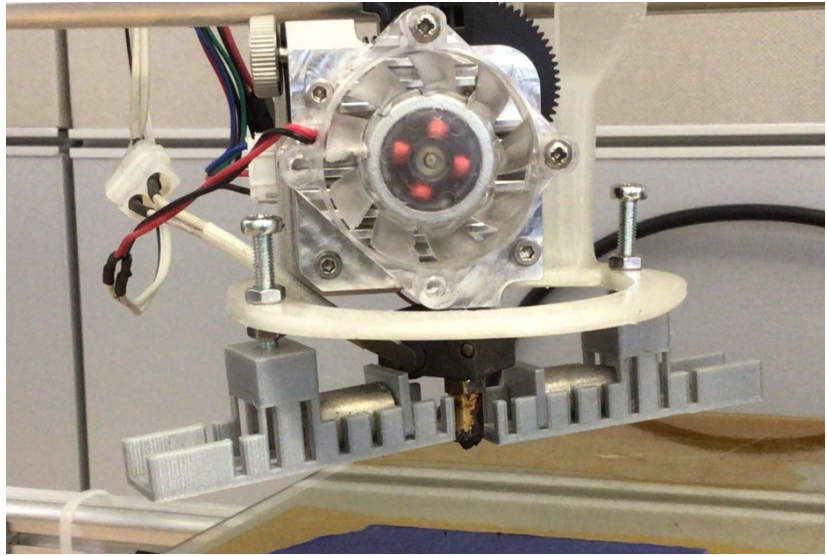

*Figure 2: Modified 3D-Printer showing the extruder including the nozzle and the flexible fixing unit for the permanent magnets*

*Table 1: Magnetic properties of the Sm₂Co₁₇ permanent magnets that are placed next to the nozzle in order to align the magnetic particles in-situ*

| | |
|---|---|
| $B_r$ | 1 100 mT |
| $H_{cJ}$ | 1194 kA/m |
| $\rho$ | 8,4 g/cm³ |
| $(BH)_{max}$ | 240 kJ/m³ |
| $T_C$ | 820°C |

The field can be varied by manually changing the distance of the permanent magnets. Due to the attraction of the printing material by the magnets, the magnets could not be placed at the minimum distance with no gap between the nozzle and the magnets. The distance has to be increased until the material remains on the printing bed during the printing process. The impact of the field was significantly different for the two investigated materials. The $Sm_2Fe_{17}N_3$ filament requires a larger distance between the magnets to avoid that the printed filament is attracted to the poles of the permanent magnets, which is due to the higher saturation magnetization compared to $SrFe_{12}O_{19}$. The magnetic field strength in the printing nozzle has been found by simulations with COMSOL Multiphysics[8]. For the simulation parameters, the magnetic properties as shown in Table 1 are used where for the $SrFe_{12}O_{19}$SrFeO, a distance of 15 mm is used and for the $Sm_2Fe_{17}N_3$, a distance of 21 mm is used (Table 2).

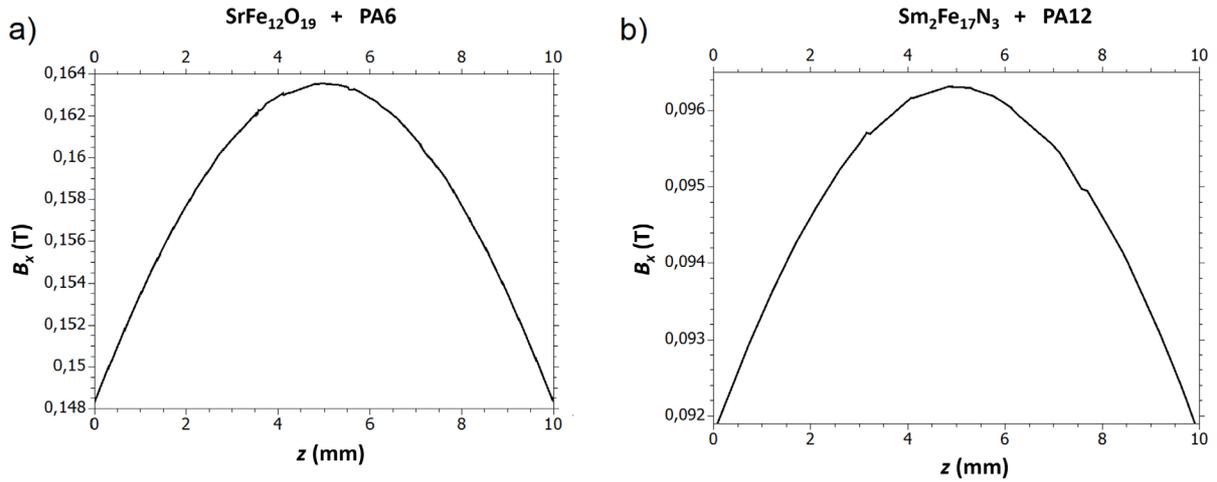

*Figure 3: Magnetic flux density $B_x$ for the hard magnetic distance of a) 15 mm which is suitable for Sprox 10/20p; b) 21 mm which is suitable for $Sm_2Fe_{17}N_3$. The zero of z=0 is the end of the nozzle which is equal to the position of the building platform*

In Fig. 3, the strength of the $B_x$ field is shown as a function of the z distance. The z coordinate is zero at the end of the nozzle, which is equal to the position of the building platform.

The settings for the manufacturing process with the 3D-printer were found empirical. In Table 2, these parameters are listed.

*Table 2: Printing parameters for the two used materials. The rectilinear filling is rotated by 90° by adjacent layers so that no flow anisotropy occurs.*

| parameter | $SrFe_{12}O_{19}$ + PA6 | $Sm_2Fe_{17}N_3$ + $PA12$ |
|---|---|---|
| distance | 15 $mm$ | 21 $mm$ |
| $B_x$ | $150 - 160$ mT | $90 - 100$ mT |
| printing temperature | 295 °C | 260 °C |
| print bed temperature | 40 °C | 60 °C |
| nozzle diameter | 0,8 mm | 0,8 mm |
| print speed | 25 mm/s | 25 mm/s |
| filling | 100% rectilinear | 100% rectilinear |
| layer height | 0,1 mm | 0,15 mm |
| filament diameter | 1,75 mm | 1,75 mm |

The fabricated samples are cubes with an edge length of 8 $mm$. The magnetic characteristics were measured with a vibrating sample magnetometer (VSM - Quantum Design PPMS-9 Tesla). For this purpose, the samples were cut into smaller cubes to be sufficiently small for the VSM (large bore option). In order to evaluate the degree of the alignment, we use the ratio of remanence to saturation magnetization $M_s$. The Stoner-Wohlfarth model of non-interacting particles predicts that an object with fully parallel aligned domains has a remanent magnetization $M_r$ that is twice as large as $M_r$ of an assembly of randomly oriented domains[9]. Hence, the ratio $M_r/M_s$ shows the degree of the alignment of the particles. If it is fully aligned, the ratio amounts to $M_r/M_s = 1$. Of course, a high degree of alignment is beneficial since it leads to higher stray fields for the applications.

In Fig. 4a, the normalized magnetization $M_r/M_s$ of the $SrFe_{12}O_{19}$ sample depending on the internal magnetic field $\mu_0 H_{int}$ is shown. $H_{int}$ is the magnetic field after the magnetization correction $H_{int} = H_d + H_{ext}$ with the demagnetization field $H_d = -N_d M$. The demagnetizing factor depends on the shape of the sample and is $N_d = 1/3$ for a cube. In Fig. 4b, the same loops for the $Sm_2Fe_{17}N_3 + PA12$ sample are shown.

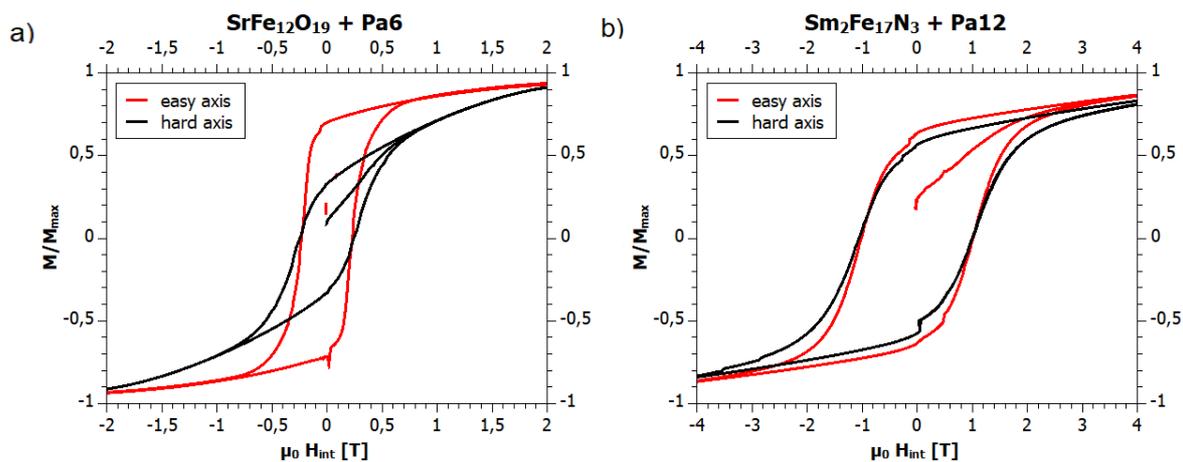

Figure 4: Hard axis and easy axis hysteresis loops of the (a) $SrFe_{12}O_{19}$ (Strontium hexaferrite) and (b) $Sm_2Fe_{14}N_3$. Here the easy axis loop denotes the direction that is parallel to the applied field during the printing process.

For $SrFe_{12}O_{19}$, the ratio $M_r/M_s$ in the direction of the easy axis is 0,70, and along the hard axis, it is 0,33. Here, we denote the direction in which the field was applied during printing as the easy axis. For $Sm_2Fe_{17}N_3 + PA12$ the normalized magnetization along the easy axis is 0,65 and 0,55 in the direction of the hard axis.

The results of the two materials are significantly different, which was expected due to the intrinsic magnetic properties of both materials. For the ferrite material, the particles could be significantly aligned. For the rare earth bonded magnet the measurements along the easy and hard axis are very similar, which implies that the field's strength was too small to align the powder along their easy axis. For the aligning of this material, a field of approximately $1\ T$, or 15 times what was possible, would be necessary according to Ref [5]. However, for large fields, we obtain the problem that the melted filament is attracted to the poles of the permanent magnet instead of the building platform.

**Conclusion**

In conclusion, this paper presents a newly designed 3D-Printer, that allows to produce anisotropic polymer-bonded magnets by aligning ferrite powder during the printing process. It is possible to print complex structures with special magnetic capabilities. It is easy to modify the construction of the fixing units of the permanent magnets so that they can be rotated and any direction of the alignment within the plane can be achieved. Hence, highly optimized magnets can be produced, such as Halbach arrays that cannot be produced by any other method.